\documentclass[prb,twocolumn,superscriptaddress]{revtex4-2}

\usepackage{graphicx,color,bm,amsmath}

\newcommand{\bfr}{{\bf r}}
\newcommand{\bfq}{{\bf q}}
\newcommand{\bfQ}{{\bf Q}}
\newcommand{\bfG}{{\bf G}}
\newcommand{\bfk}{{\bf k}}

\begin{document}

\title{Finite-momentum dielectric function and excitonic effects from time-dependent density-functional theory with dielectrically screened hybrid functionals}
\author{Didarul Alam}
\affiliation{Department of Physics and Astronomy, University of Missouri, Columbia, Missouri 65211, USA}
\author{Jiuyu Sun}
\email{sunjiuyu@njust.edu.cn}
\affiliation{Department of Applied Physics and MIIT Key Laboratory of Semiconductor Microstructure and Quantum Sensing, Nanjing University of Science and Technology, Nanjing 210094, China}
\author{Carsten A. Ullrich}
\affiliation{Department of Physics and Astronomy, University of Missouri, Columbia, Missouri 65211, USA}

\begin{abstract}
This paper studies the performance of time-dependent density-functional theory (TDDFT) for calculating the dielectric function of semiconductors and insulators
at finite momentum transfer, comparing against the standard Bethe-Salpeter equation (BSE). Specifically, we consider a recently proposed hybrid approach that
mixes dielectrically screened exact exchange with a semilocal functional, and we also introduce a new hybrid functional featuring a truncated dielectric screening scheme.
The computational effort of these hybrid TDDFT approaches is significantly
less than that of the BSE, but they deliver comparable accuracy, as demonstrated for the semiconductors Si and GaN and the wide-band insulator LiF.
This opens up possibilities for calculating exciton dispersions and electron energy loss functions efficiently and accurately for a wide range of materials.
\end{abstract}

\maketitle

\section{Introduction} \label{Introduction}
Excitonic effects are critical for the optical properties of electronic materials in light-emitting devices, for photovoltaics, and for  photocatalysts \cite{wheeler2013exciton, anantharaman2021exciton,dresselhaus2007exciton}. Excitons are bound quasiparticle states of electrons and holes generated during excitation processes close to the band
gap; they are essential for light-matter interaction processes, and they have given rise to a growing research area called quantum excitonics \cite{Hu2024}.
The standard theoretical description of excitons is via the Bethe-Salpeter equation (BSE) \cite{sham1966many, hanke1979many,hanke1980many,albrecht1998ab,Rohlfing1998, rohlfing2000electron,onida2002electronic}, a Green’s-function based approach known for its accuracy in calculating optical properties of insulators and semiconductors.
The BSE explicitly accounts for the screened Coulomb interactions within electron-hole pairs, including direct and exchange contributions. However, this accuracy comes with a relatively high computational cost, which largely limits its application to a wider range of materials of interest.

The BSE framework involves several key steps. Starting from the many-body Hamiltonian, it transitions to Green's functions to handle particle interactions. The GW approximation \cite{PhysRev.139.A796,aryasetiawan1998gw} is then used to calculate the self-energy, essential for formulating the BSE. The dielectric matrix, playing a central role for accounting for screening of the Coulomb interaction, is typically computed using the Random Phase Approximation (RPA) \cite{PhysRev.82.625,PhysRev.126.413}. Solving the BSE then involves building the electron-hole interaction kernel and diagonalizing the resulting large matrix to obtain exciton binding energies and optical spectra. All of this can be computationally quite demanding; consequently, there is a need for alternative methodologies that can maintain accuracy while reducing computational effort.

Within the framework of time-dependent density-functional theory (TDDFT), Sun {\em et al.} recently proposed an alternative to the BSE approach focusing on simplifying the screened Coulomb interaction \cite{Sun2020low}. This study highlights the growing potential of hybrid functionals \cite{becke1993new, perdew1996rationale} in materials science due to their ability to accurately approximate quasiparticle band structures and improve band gap predictions \cite{Jiang2018}. Over the past decade, hybrid exchange-correlation (XC) functionals, which combine a fraction of nonlocal (Hartree-Fock) exchange with semilocal exchange and correlation, have become popular in ground-state DFT for solving the band-gap problem \cite{Matsushita2011, Jain2011, henderson2011accurate,Chen2018} and have also shown promise for describing excitations in periodic solids \cite{PhysRevB.78.121201, PhysRevB.83.195325, webster2015optical, hehn2022excited}. Recent studies have shown good agreement with GW/BSE results for a range of materials \cite{PhysRevMaterials.3.064603, PhysRevMaterials.3.084007,Maji2022,Ohad2022,Ohad2023,Camarasa2023,Ghosh2024}.

In a similar context, a screened exact exchange (SXX) approach, proposed earlier by Yang {\em et al.} \cite{Yang2015screened}, can be viewed as a simplified BSE method in which the dielectric function of three-dimensional (3D) bulk materials is replaced by a uniform screening parameter $\gamma$. This parameter, determined non-empirically as the inverse of the dielectric constant, requires RPA calculations only at wavevector $\bfq \rightarrow 0$, significantly reducing computational demand; it can also be obtained from experimental dielectric constants. The dielectric function can thus be approximated using simple models or a single screening parameter $\gamma$ without significant loss of accuracy. Integrating this simplified BSE kernel with a local XC kernel within a hybrid functional framework effectively captures exciton binding energies and optical spectra across various materials, presenting a non-empirical and efficient alternative to the full BSE method \cite{Sun2020low,Byun_2020,Sun2020b}. In addition, the single
screening parameter can be made more sophisticated by adopting an adjustable range-separated screening \cite{Lewis_Sahar2020,Chen2018,Alexey2020}, showing a transferable performance in more materials \cite{Wing2019,PhysRevMaterials.3.084007,Camarasa2023}. Thus, this hybrid framework combines the strengths of both worlds --- many-body perturbation theory and (TD)DFT --- to achieve precise results with reduced computational demands \cite{Garrick_Kronik2020,HongJiang2020}.

The prior studies discussed above mainly focused on determining optical absorption spectra in the long-wavelength ($\bfQ \rightarrow 0$) limit.
There is, however, considerable interest in excitation processes occurring at finite momentum transfer, which are experimentally probed using
scattering techniques including inelastic X-ray scattering (IXS) or electron energy loss spectroscopy (EELS) \cite{Nicolau2025}.
Several theoretical and computational studies have developed GW/BSE and TDDFT methodologies to calculate IXS or EELS spectra \cite{Weissker2006,botti2007time, Weissker2010, sharma2012enhanced} as well as exciton dispersions and band structures \cite{gatti2013exciton, Fugallo2015, Cudazzo2016, Deilmann2019, Fugallo2021, Haber2023, Qiu2021, Mei2022, Liu2023}.
Interestingly, the adiabatic local density approximation (ALDA), which fails for excitons, was found to accurately treat the dielectric tensor and EELS spectra at finite
values of  $\bfQ$  \cite{Weissker2006,botti2007time, Weissker2010}. This raises the question about the performance of hybrid TDDFT for
finite values of $\bfQ$.

Thus, in this paper, we extend the analysis to the calculation of dielectric functions at finite momentum transfer. In addition to our dielectrically screened hybrid
functional from Ref. \cite{Sun2020low} with a uniform, global screening parameter, we introduce a new hybrid functional featuring a momentum-dependent but
truncated dielectric screening, which can be viewed as a simplified range-dependent hybrid functional. We apply our hybrid TDDFT approaches to
the 3D materials silicon (Si), lithium fluoride (LiF), and wurtzite gallium nitride (GaN), which are chosen for their diverse electronic properties and significance in fundamental research and technological applications. We shall demonstrate that both optical spectra and exciton dispersions can be accurately reproduced, closely matching results from full BSE calculations. Moreover, we also highlight the importance of range-separated screening for constructing the hybrid functionals,
aiming to be compatible with a wider range of materials.

The paper is organized as follows. In Sec. \ref{Theoretical} we present the theoretical background of the BSE and hybrid TDDFT formalisms and in Sec. \ref{Comp_Details} we discuss some computational details. In Sec. \ref{Result} we present results for insulators and semiconductors using both BSE and hybrid TDDFT for finite $\bfQ$. Sec. \ref{Summary} contains conclusions and outlook.

\section{Theoretical background: full BSE, simplified BSE, and hybrid TDDFT} \label{Theoretical}

\subsection{BSE, model-BSE, and SXX}

We begin by briefly reviewing the calculation of optical properties in solids based on generalized TDDFT with hybrid functionals \cite{byun2020time, Sun2020low}. A direct connection exists between the BSE and generalized TDDFT, since BSE can be formulated as a matrix equivalent to the Casida equation in TDDFT \cite{casida1998molecular,ullrich2011time}. By employing the Tamm-Dancoff approximation \cite{hirata1999time, sander2015beyond}, one obtains the BSE with finite momentum transfer $\bfQ$ \cite{Qiu2015}:
\begin{eqnarray} \label{eq:1}
\lefteqn{\hspace{-3cm}
\sum_{v'c'\bfk'}\left[(E_{c\bfk + \bfQ} - E_{v\bfk}) \delta_{vv'}\delta_{cc'}\delta_{\bfk\bfk'} +K^{\rm BSE,\bfQ}_{cv\bfk,c'v'\bfk'} \right]Y^{n\bfQ}_{v'c'\bfk'}  } \nonumber\\
&=& \omega_{n\bfQ} Y^{n\bfQ}_{vc\bfk} \:.
\end{eqnarray}
Here, $v$ and $c$ denote occupied valence and unoccupied conduction bands, respectively, and $E_{c\bfk}$ and $E_{v\bfk}$ are the associated energy dispersions.
The  BSE kernel or coupling matrix
$K^{\rm BSE}$ is usually expressed as the sum of a direct and exchange electron-hole kernel \cite{Patterson2005,Fuchs2008,Bieniek2022}:
\begin{equation} \label{eq:2}
    K^{\rm BSE} = K^{d} + K^{x}.
\end{equation}
Here, the (repulsive) exchange term is given by
\begin{eqnarray} \label{eq:3}
K^{x,\bfQ}_{cv\bfk,c'v'\bfk'} &=& \frac{2}{V_{\rm cell}} \sum_{\bfG \neq 0} \frac{4\pi}{|\bfG + \bfQ|^2} \langle c\bfk+\bfQ | e^{i(\bfG+\bfQ) \cdot \bfr} | v\bfk \rangle \nonumber\\
&\times&
\langle v'\bfk' | e^{-i(\bfG+\bfQ) \cdot \bfr} | c'\bfk'+\bfQ \rangle,
\end{eqnarray}
where $V_{\rm cell}$ is the unit cell volume, and the (attractive) direct electron-hole interaction term is
\begin{eqnarray} \label{eq:4}
\lefteqn{K^{d,\bfQ}_{cv\bfk,c'v'\bfk'} =
-\frac{2}{V_{\rm cell}} \sum_{\bfG, \bfG'} W_{\bfG \bfG'}(\bfq,\omega) \delta_{\bfq, \bfk - \bfk'}} \\
&\times&
\langle v'\bfk'| e^{i(\bfq + \bfG) \cdot \bfr} | v \bfk \rangle \langle c\bfk+\bfQ  | e^{-i(\bfq + \bfG') \cdot \bfr} | c'\bfk'+\bfQ \rangle. \nonumber
\end{eqnarray}
In our hybrid TDDFT calculation $K^x$ will remain unchanged.  The direct term, $K^d$, is much more computationally expensive due to the screened Coulomb interaction,
\begin{equation} \label{eq:5}
W_{\mathbf{G} \mathbf{G}'} (\bfq, \omega) = \frac{4\pi \varepsilon_{\bfG \bfG'}^{-1} (\bfq, \omega)}{|\bfq + \bfG| |\mathbf{q} + \mathbf{G}'|},
\end{equation}
which features the inverse dielectric matrix
\begin{equation} \label{eq:6}
    \varepsilon_{\bfG\bfG'}^{-1} (\mathbf{q}, \omega) = \delta_{\bfG, \bfG'} + \frac{4\pi}{|\mathbf{q} + \mathbf{G}|^2} \chi_{\bfG\bfG'}^{\rm RPA} (\mathbf{q}, \omega),
\end{equation}
typically calculated in RPA. Note here the $\bfQ$ is the finite exciton center-of-mass momentum, which is set to zero when calculating vertical excitations in most BSE codes.
In the following part of this section, we mainly show the formalism in the $\bfQ \rightarrow 0$ limit for simplicity; it can easily be generalized for the finite $\bfQ$ form.

In most applications of the BSE, the dielectric matrix is used in its static limit, $\varepsilon^{-1}_{\bfG,\bfG'}(\bfq, \omega = 0)$ \cite{Rohlfing1998, benedict1998optical}. This calculation can be computationally costly, as it involves summing over the $\bfq$-grid and performing a double sum over $\bfG$ and $\bfG'$. Furthermore, achieving full convergence for $\varepsilon^{-1}$ usually necessitates the inclusion of many unoccupied bands.
If the BSE follows a GW calculation for the band structure, then the dielectric matrix is usually already available. However, if the band structure is
calculated using other methods (for instance DFT plus scissors, as done here), then the dielectric matrix can become a true bottleneck.

If we were to neglect dielectric screening altogether by setting $\varepsilon^{-1} = 1$ in Eq. (\ref{eq:5}), then the BSE would reduce to the time-dependent Hartree-Fock (TDHF) method \cite{dreuw2005single}. Unscreened TDHF is notorious for causing severe overbinding of excitons (except in cases of extremely strongly bound excitons such as in solid Ar or Ne), potentially leading to a complete collapse of the spectral structure.

A practical compromise between the full dielectric screening of BSE and the no-screening TDHF approach is to adopt a simplified dielectric screening. Yang {\em et al.} \cite{Yang2015screened} proposed their SXX method, where the dielectric matrix is replaced by a single screening parameter $\gamma$, which is diagonal in $\bfG$ and $\bfG'$:
\begin{equation} \label{eq:7}
\varepsilon^{-1}_{\bfG\bfG'}(\bfq) \rightarrow \gamma \delta_{\bfG\bfG'}.
\end{equation}
The screening parameter $\gamma$ can be calculated from first principles as $\gamma = (\varepsilon^{\rm RPA})^{-1}_{00} (\bfq = 0)$.
A somewhat more sophisticated approach,
\begin{equation} \label{eq:8}
    \varepsilon^{-1}_{\bfG\bfG'}(\bfq) \rightarrow \varepsilon_m(q)^{-1} \delta_{\bfG\bfG'},
\end{equation}
replaces the single parameter $\gamma$ by a model dielectric function
$\varepsilon_m(q)$ \cite{Cappellini1993,Lorin2021}. In the following we refer to the latter method as m-BSE; the model dielectric function used as input is the one proposed for 3D (bulk) systems by Cappellini {\em et al.} \cite{Cappellini1993}:
\begin{equation} \label{eq:m_BSE}
   \varepsilon_m(q) = 1 + \left[\frac{1}{\epsilon(0) - 1} + \alpha \left(\frac{q}{q_{\text{TF}}}\right)^2 + \frac{\hbar^2 q^4}{4m^2 \omega_p^2}\right]^{-1},
\end{equation}
where $\epsilon(0)$ is the static dielectric constant (obtained from experiment or RPA calculations),
$\alpha = 1.563$ is an empirical parameter, $q_{\text{TF}}$ is the Thomas-Fermi wave vector related to the average electron density, and $\omega_p$ is the plasma frequency of the material. This replacement avoids the computation of the full $\epsilon^{-1}_{\bfG\bfG'}(\bfq)$, significantly reducing computational cost.
A comparison \cite{Sun2020low} showed that both SXX and m-BSE, Eqs. (\ref{eq:7}) and (\ref{eq:8}),
produced exciton binding energies and optical spectra of similar quality as and in good agreement with full BSE calculations.

\subsection{Hybrid TDDFT: global and truncated screening}

Over the past decade, hybrid XC functionals have been effectively applied to describe excitations in periodic solids. Sun {\em et al.} \cite{Sun2020low} introduced a nonempirical hybrid approach specifically designed to accurately reproduce excitonic properties in close agreement with those obtained from the BSE. This method employs the strengths of both the long-range SXX and ALDA to provide a balanced treatment of XC effects.
The core idea is to combine the accurate long-range behavior of the SXX with an approximate but efficient treatment of local-field effects provided by the ALDA. This combination aims to retain the essential electron-hole interaction physics captured by the BSE while significantly reducing computational complexity.

In practical terms one achieves this by modifying the BSE kernel, Eq. (\ref{eq:2}) by replacing the direct kernel $K^d$ with a hybrid kernel that incorporates both the SXX and ALDA components:
\begin{equation} \label{eq:10}
    K_{\rm xc}^{\rm gDDH} = K^{\rm SXX} + (1 - \gamma) K_{\rm xc}^{\rm ALDA} \:,
\end{equation}
where gDDH denotes global dielectric-dependent hybrid.
Here,  $K^{\rm SXX}$ and $K_{\rm xc}^{\rm ALDA}$ are expressed as follows:
\begin{eqnarray} \label{eq:11}
K^{\rm SXX}  &=& -\frac{2}{V_{\rm cell}} \sum_{\bfG} \left[ \delta_{\bfq, \bfk - \bfk'} \frac{4 \pi \gamma}{|\mathbf{q} + \mathbf{G}|^2} \right] \\
&\times&
\langle c \mathbf{k} | e^{i (\mathbf{q} + \mathbf{G}) \cdot \mathbf{r}} | c' \mathbf{k'} \rangle \langle v' \mathbf{k'} | e^{-i (\mathbf{q} + \mathbf{G}) \cdot \mathbf{r}} | v \mathbf{k} \rangle, \nonumber
\end{eqnarray}
and
\begin{eqnarray} \label{eq:12}
K_{\rm xc}^{\rm ALDA}  &=& \frac{2}{V_{\rm cell}} \lim_{\bfq \to 0} \sum_{\bfG, \bfG'} f_{{\rm xc}, \bfG \bfG'}^{\rm ALDA} (\mathbf{q}) \\
&\times& \langle c \mathbf{k} | e^{i (\mathbf{q} + \mathbf{G}) \cdot \mathbf{r}} | v\mathbf{k} \rangle \langle v' \mathbf{k'} | e^{-i (\mathbf{q} + \mathbf{G}) \cdot \mathbf{r}} | c' \mathbf{k'} \rangle. \nonumber
\end{eqnarray}
The gDDH kernel is designed to capture both the long-range interactions and the local-field effects that are crucial for accurately describing excitonic properties.
This approach reproduces excitonic binding energies with high accuracy, closely matching the results from full BSE calculations, while also benefiting from a reduced computational cost \cite{Sun2020low,Sun2020b}.

\begin{figure}
    \includegraphics[width=\linewidth]{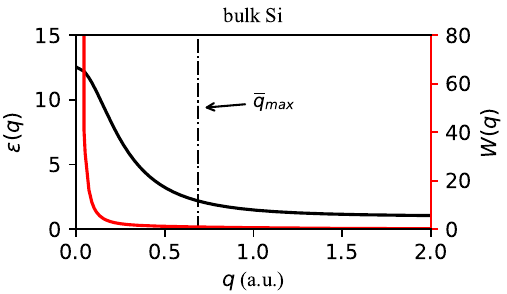}
    \caption{The $q$-dependent dielectric function $\varepsilon(q)$ (left axis in black) and the screened Coulomb potential (arbitrary units) in Eq. (\ref{eq:5}) (right axis in red) of bulk Si, calculated by Eq. (\ref{eq:m_BSE}) with the head term $\varepsilon(0)=(\varepsilon^{\rm RPA})^{-1}_{00} (\bfq = 0)$. The vertical dashed line indicates the $\bfq$ with the largest module in the first Brillouin zone ($\Bar{\bfq}_{max}$).}
    \label{fig:Si_eps}
\end{figure}

However, the gDDH functional presented above does not always deliver an optimal agreement with standard BSE for the optical spectra of semiconductors like bulk Si; the m-BSE usually has a better performance in most cases \cite{Sun2020low}. To construct an improved hybrid functional, let us consider
Fig.~\ref{fig:Si_eps}, which shows the $q$-dependent dielectric function for bulk Si: as can be seen, $\varepsilon(q)$ starts out at 13 at $q=0$ and approaches 1 in the $q \rightarrow \infty$ limit. Thus, the use of a uniform parameter $\gamma = (\varepsilon^{\rm RPA})^{-1}_{00} (\bfq = 0)$ in  gDDH overestimates the screening effect. On the other hand, as shown in Fig.~\ref{fig:Si_eps}, the screened Coulomb potential $W(q)$ rapidly decreases with $q$ and becomes almost negligible
once $q$ goes beyond 0.5 a.u. To account for these properties, we now propose a truncated dielectric-dependent hybrid (tDDH) functional, with the
momentum-dependent screening parameter
\begin{equation}\label{13}
 \gamma = \varepsilon(\Bar{\bfq})^{-1},
\end{equation}
where the $\Bar{\bfq}$ stands for the $\bfq$-vectors confined in the first Brillouin zone ($\bfG=\bfG'=0$). In practice, $\varepsilon(\Bar{\bfq})^{-1}$ can be obtained via RPA at each individual $\bfq$, or by combining the head of $(\varepsilon^{\rm RPA})^{-1}_{00} (\bfq = 0)$ from RPA and a $\bfq$-dependent model dielectric function such
as Eq. (\ref{eq:m_BSE}).
We point out that tDDH can be viewed as an approximation of the range-separated DDHs, in which the error function $\mbox{erf}(\gamma|\bfr-\bfr'|)$ offers a good model for the dielectric function $\varepsilon^{-1}(\bfr,\bfr')$ in 3D solids \cite{Skone2016,Chen2018}.
Our $q-$truncated screening scheme, by dropping off the tail ($|\bfG|\neq0$ and off-diagonal ($\bfG \neq \bfG'$) terms in the dielectric matrix,
not only saves the computational cost of the full RPA and avoids potential numerical instability of the ALDA kernel,
but also takes the essentials of dielectrically screened exchange interaction in BSE as we will demonstrate in the following.

\section{Computational Details} \label{Comp_Details}

Our goal is to assess the performance of the TDDFT hybrid kernels compared against the BSE, using the same ground-state band structures as input;
for this purpose, simple DFT (rather than expensive GW) band structures are sufficient. We consider three materials:
bulk Si and two commonly studied binary compounds, the semiconductors GaN and the wide-bandgap insulator LiF.
To compute the Kohn-Sham band structures, we use the LDA functional within the Quantum Espresso package \cite{giannozzi2017advanced}, utilizing a plane-wave basis and optimized norm-conserving Vanderbilt pseudopotentials \cite{ONCVPSP1,ONCVPSP2}. The electronic band gaps were adjusted to match the experimental values using scissors operators \cite{PhysRevLett.63.1719}, with experimental lattice parameters applied to all materials.

We used the Yambo code \cite{marini2009yambo,Yambo2019} for calculating the macroscopic dielectric function for finite momentum transfer, $\varepsilon(\bfQ,\omega)$, using a customized implementation of the TDDFT hybrid functionals. We used $24 \times 24 \times 24$, $16 \times 16 \times 16$ and $16 \times 16 \times 10$ $k$-point meshes for Si, LiF and GaN, respectively. The RPA and ALDA spectra are calculated within the linear-response theory framework by solving a Dyson-like equation \cite{ullrich2011time}. The response functions ($\chi$) are computed with 100 bands and we included $\bfG$-vectors up to 200 reciprocal lattice vectors. For constructing the BSE, m-BSE, and hybrid kernels we utilized 4 valence and 4 conduction bands for Si and LiF, and 6 valence and 6 conduction bands for GaN. To obtain the optical spectra we adapted the calculations by using Haydock iteration \cite{haydock1980recursive} to solve the BSE-type equations, rather than directly diagonalizing the large BSE matrix. The spectra are presented with a broadening parameter of 0.1 eV to simulate typical experimental linewidths. For the hybrids we adopted $\gamma=\varepsilon^{-1}_{00} (\bfq = 0)$ from RPA for gDDH, and subsequently utilize the model dielectric function in Eq. (\ref{eq:m_BSE}) for tDDH.

\section{Results: dielectric function at finite momentum transfer} \label{Result}

\subsection{Silicon (Si)}

Bulk Si has a face-centered cubic (FCC) lattice structure featuring two atoms per unit cell, and an experimental lattice constant of $a=10.2$ a.u. With LDA we obtain a direct band of gap 2.57 eV at the $\Gamma$ point and an indirect band gap of 0.54 eV. Using the scissors operator we enforce the correct indirect gap of 1.12 eV.

\begin{figure*}[htp]
\centering
    \includegraphics[width=\linewidth]{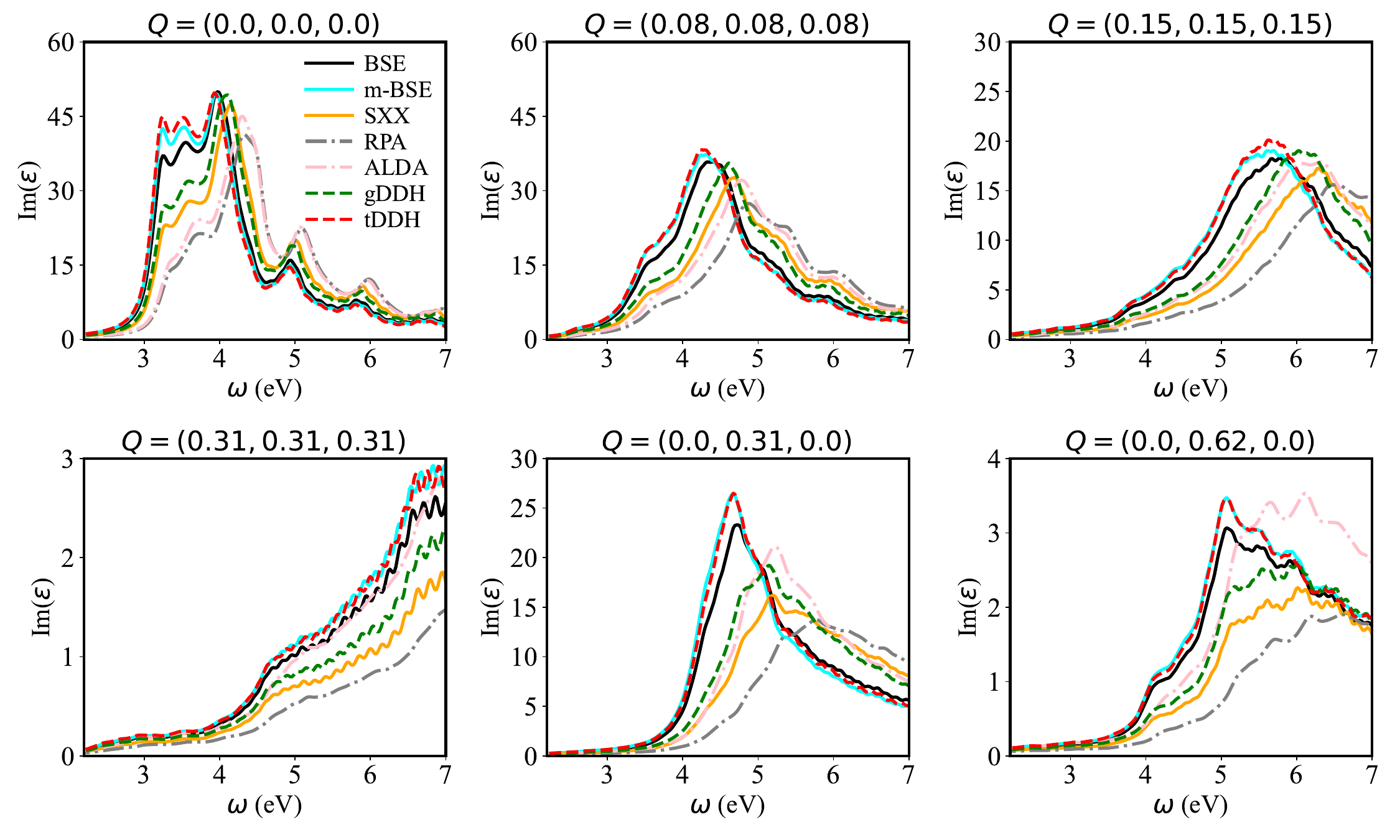}
    \caption{Imaginary part of the dielectric function $ {\rm Im}(\varepsilon(\bfQ,\omega))$ of Si, calculated using BSE, m-BSE, gDDH, tDDH, RPA, and ALDA for various
    momentum transfers: (a) $\bfQ = (0,0,0)$;  (b) $\bfQ = (0.08, 0.08, 0.08)$; (c) $\bfQ = (0.15, 0.15, 0.15)$; (d) $\bfQ = (0.31, 0.31, 0.31)$; (e) $\bfQ = (0.0, 0.31, 0.0)$; (f) $\bfQ = (0.0, 0.62, 0.0)$. All Cartesian $\bfQ$-components are given in a.u.}
    \label{fig:Si_q}
\end{figure*}

Figure \ref{fig:Si_q} illustrates the imaginary part of the dielectric function ${\rm Im}(\varepsilon(\bfQ,\omega))$ for Si, calculated using five different methods: BSE, m-BSE, gDDH, tDDH, RPA (which ignores xc effects in the linear response altogether), and ALDA. The dielectric functions are computed for a range of momentum transfer $\bfQ$ along [111] and [010] direction in reciprocal space in a.u. (see figure caption for more details about the individual components of each $\bfQ$).
We note that the FCC Brillouin zone is $4\pi/a$ across, so the reciprocal-space distance between the $\Gamma$ and $X$-points in Si is 0.612 a.u. and the momentum transfers
considered here are all within one reciprocal lattice vector.

At $Q = 0.00$ a.u., which defines the vertical (optical) absorption spectrum, the BSE method captures the excitonic effects and shows a prominent excitonic peak around 3.2 eV,
as well as a strong peak at 4.2 eV due to a van Hove singularity.
The ALDA and RPA fail to account for excitonic effects, as is well documented in the literature \cite{onida2002electronic}.
As in previous work \cite{Sun2020low}, the introduction of exact-exchange interaction makes the SXX and gDDH able to improve the description of the peak at 3.2 eV against the ALDA, which is however still underestimated compared to BSE, due to the overscreening by a single macroscopic dielectric constant.

On the other hand, the m-BSE and tDDH slightly overestimate the peak at 3.2 eV, because both of them neglect the off-diagonal terms in the $K^d$ kernel (due to
$\delta_{\bfG \bfG'}$). Nevertheless, m-BSE and tDDH show better performance for the overall optical spectrum against their global screened counterparts (SXX and DDH), which indicates the importance of range-separated screening when describing electron-hole interaction in this optical regime.

As $Q$ increases, the spectra become broader and the peaks shift towards higher energies across all methods, leading to a redistribution of oscillator strength
to form a broad maximum in the 4-7 eV region. This behavior reflects the reduced excitonic effects at higher momentum transfers, as can be clearly seen from Fig. \ref{fig:Si_q} when comparing from $Q = 0.00$ to $\bfQ = (0.31, 0.31, 0.31)$ along [111]. These trends are generally reproduced by all five methods.

For finite momentum transfers, our results show that the ALDA kernel gives a decent account of the shape of the spectra for most finite $\bfQ$, and
even perfectly matches the BSE at $\bfQ = (0.31, 0.31, 0.31)$.
This is not surprising: the ALDA is known to correctly capture the short-range many-body effects that are crucial for a correct description of $\varepsilon(\bfQ,\omega)$ in silicon at finite momentum transfer \cite{Weissker2010}. This is not the case with the RPA, which compares only poorly with the BSE in this regime.

The important role of the short-range many-body effects can also be observed by comparing the results from SXX and gDDH. We see that the gDDH spectra are much closer to those of BSE, especially for the region of 4-6 eV, which is caused the absence of ALDA contribution in SXX.
Moreover, the tDDH reproduces the spectra by BSE best, exhibiting a noticeable improvement to the gDDH, meaning the range-separated screening is also important for the finite-$\bfQ$ excitonic effects in the semiconductors with large macroscopic dielectric constant.

\subsection{Gallium Nitride (GaN)}

We next consider GaN as a material that is in between Si and LiF in terms of the size of the band gap and macroscopic dielectric constant, as well as the strength of excitonic effects.
GaN is a wide band-gap semiconductor whose equilibrium crystal structure is the hexagonal close-packed (HCP) wurtzite form ($w$-GaN), with experimental lattice constants
$a = 6.03$ a.u. and $c = 9.80$ a.u. The reciprocal-space distances between the $\Gamma$ and $A$-points and between the $\Gamma$ and $M$ points are
0.32 a.u. and 0.35 a.u., respectively. We obtain an LDA band gap of 1.94 eV at the $\Gamma$ point and use a scissors shift of 1.46 eV to match the experimental band gap 3.4 eV \cite{levinshtein2001properties}.

\begin{figure*}[htb]
\centering
    \includegraphics[width=\linewidth]{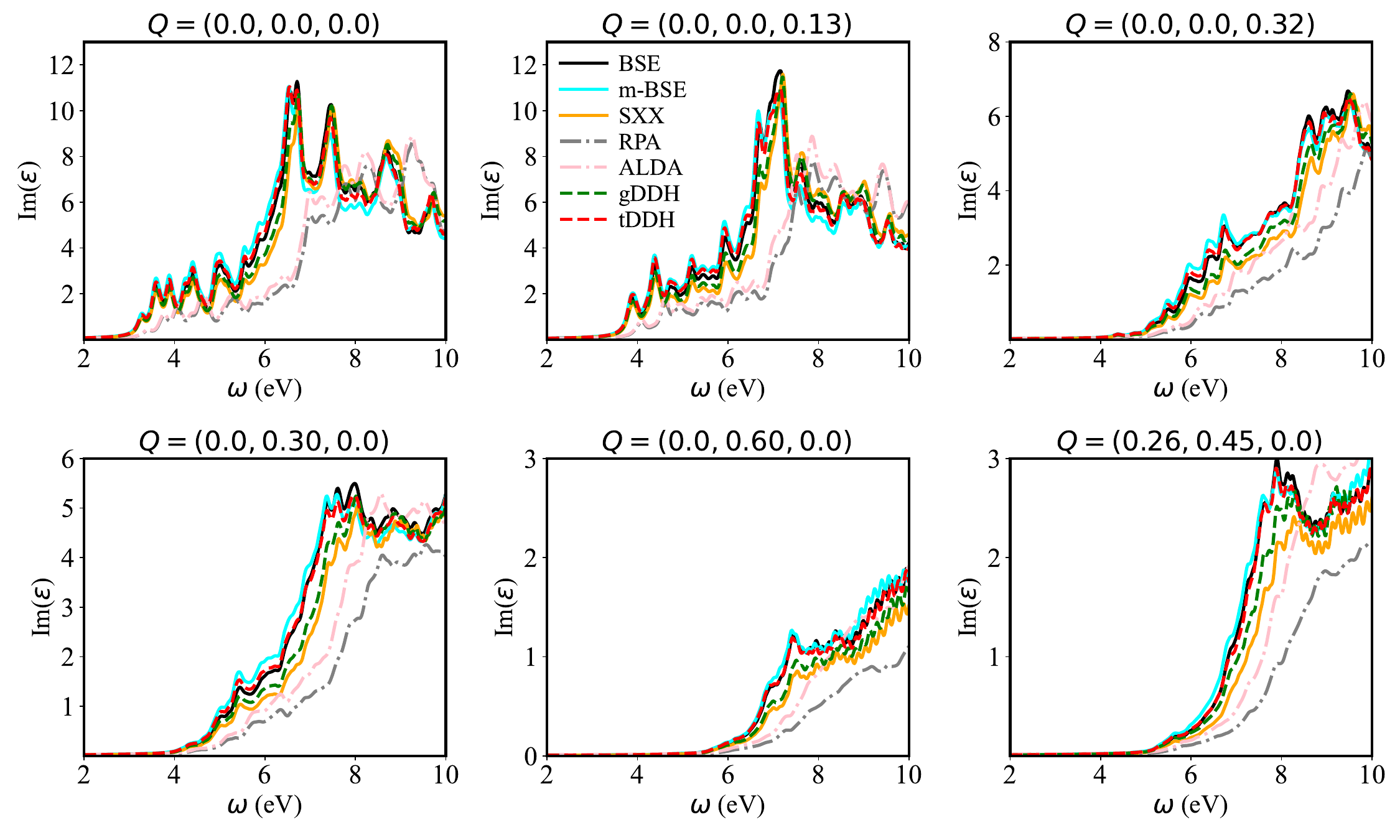}
    \caption{Same as Fig. \ref{fig:Si_q}, but for the wurtzite semiconductor GaN. The momentum transfers are as follows: a) $\bfQ = (0,0,0)$;  (b) $\bfQ = (0.0, 0.0, 0.13)$;
   (c) $\bfQ = (0.0, 0.0, 0.32)$; (d) $\bfQ = (0.0, 0.30, 0.0)$; (e) $\bfQ = (0.0, 0.60, 0.0)$; (f) $\bfQ = (0.26, 0.45, 0.0)$.}
    \label{fig:GaN_q}
\end{figure*}

In Fig. \ref{fig:GaN_q} we plot $\rm Im(\varepsilon(\bfQ,\omega))$ of GaN using the same methods as in bulk Si, considering six different $\bfQ$ values. At $Q = 0.00$ a.u., the spectra by BSE and m-BSE and the tDDH are very close to each other, and the ones by SXX and gDDH are slightly lower.
In contrast, the spectral strength in ALDA and RPA is shifted towards higher energies, failing to give the pronounced peaks above the band gap in the 6-8 eV energy range. Below the band gap at 3.4 eV, the excitonic enhancement is clearly visible in BSE and hybrids, and missing in ALDA and RPA.

As the momentum transfer $Q$ increases, the spectra undergo significant changes: the peak intensities and positions shift towards higher frequencies, and
the excitonic enhancement close to the band gap fades away.
We notice similar trends as for the case of silicon, namely, at increasing $Q$ the spectra computed with
the ALDA and with the hybrid functionals become more and more similar to each other, and both become more and more similar to the BSE.
However, ALDA here in GaN is no longer able to reproduce the spectrum by BSE at large $Q$ as for the case of Si. This may be attributed to an enhancement of long-range excitonic effects as the dielectric screening is getting weaker, which will be more significant in the wide-gap insulator with small dielectric constants (this will be even more
significant for LiF, see the following section).

One the other hand, the m-BSE, SXX, gDDH and tDDH spectra stay close to the BSE spectra, which again emphasizes the reliable performance of the kernels including screened exact-exchange interaction when describing excitonic effects.
Notably, for all the spectra with different $\bfQ$, the improvement by tDDH against gDDH is less significant than that in bulk Si, which implies the redudec importance of range-separated screening as the macroscopic dielectric constant becomes smaller.

\subsection{Lithium Fluoride  (LiF)}

\begin{figure*}[htb]
\centering
   \includegraphics[width=\linewidth]{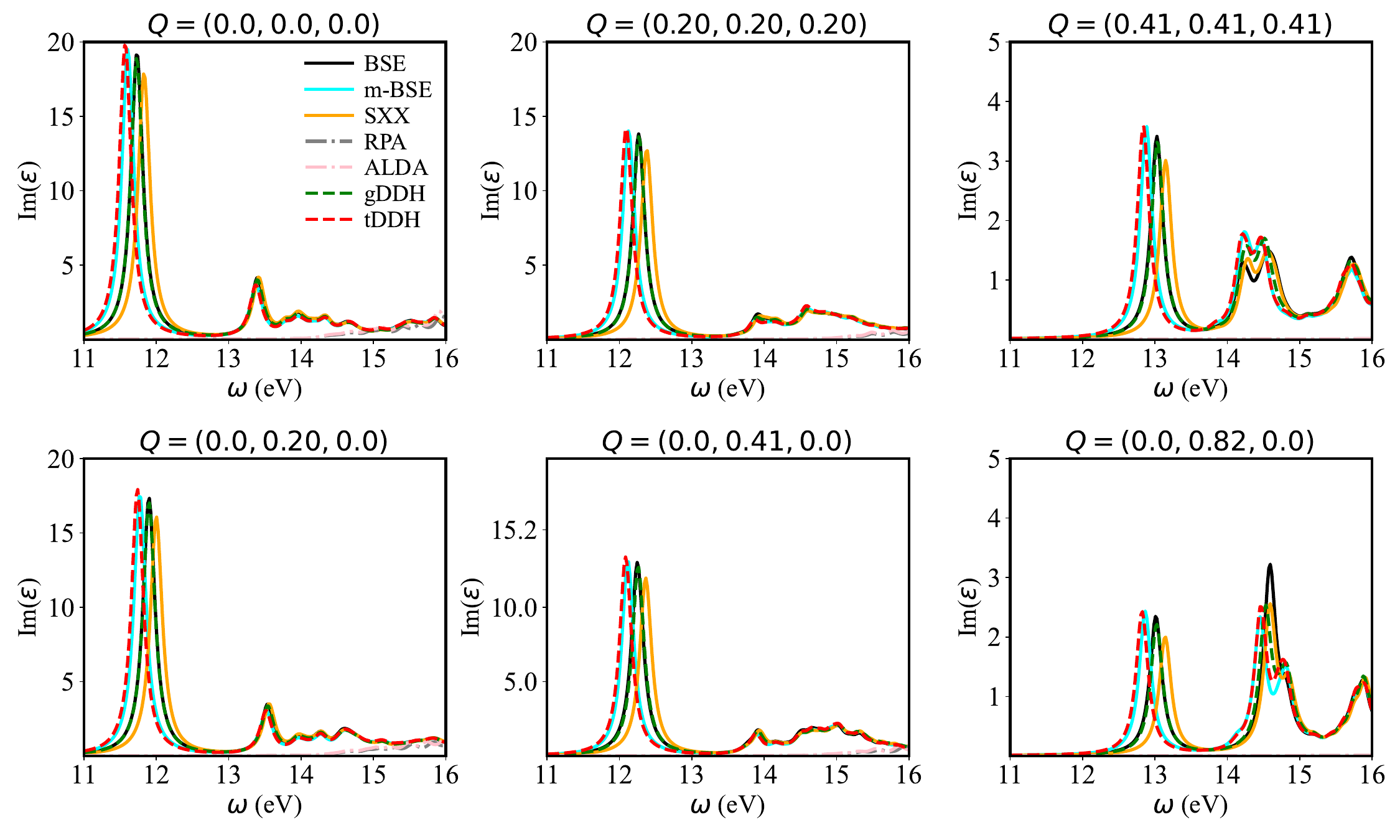}
   \caption{Same as Fig. \ref{fig:Si_q}, but for the wide-gap insulator LiF. The momentum transfers are as follows: (a) $\bfQ = (0,0,0)$;  (b) $\bfQ = (0.20, 0.20, 0.20)$;
   (c) $\bfQ = (0.41, 0.41, 0.41)$; (d) $\bfQ = (0.0, 0.20, 0.0)$; (e) $\bfQ = (0.0, 0.41, 0.0)$; (f) $\bfQ = (0.0, 0.82, 0.0)$.}
    \label{fig:LiF_q}
\end{figure*}

LiF is a wide band-gap insulator with an FCC lattice structure and lattice constant $7.63$ a.u., leading to a reciprocal-space distance between the $\Gamma$ and $X$-points
of 0.82 a.u. We obtain a band gap of 8.45 eV at the $\Gamma$ point using LDA and apply a scissors shift of 5.15 eV to reproduce the experimental gap of 13.6 eV \cite{roessler1967electronic}. The optical spectrum of LiF is notably dominated by a strongly bound exciton with a binding energy of approximately 3 eV
\cite{Rohlfing1998,Marini2003}, a feature that traditional independent particle approximations, as well as ALDA and RPA, fail to capture, resulting in a poor description of the experimentally observed spectral structures \cite{Sun2020low}.

Figure \ref{fig:LiF_q} illustrates the imaginary part of the dielectric function of LiF at six different momentum transfer values, ranging from $Q$ = $0.00$ a.u. to the large $Q$ = $0.82$ a.u. at the edge of the first Brillouin zone. At $Q = 0.00$ a.u., the BSE method exhibits a series of pronounced excitonic peaks, with the lowest peak at about 11.9 eV and a second excitonic peak at around 13.4 eV. The ALDA and RPA spectra are almost invisible in the figures, showing again the strong excitonic effects in large gap  LiF.
As for the case of Si, m-BSE and tDDH generate spectra that are very close to the BSE, with only slight deviations due to the underscreened Coulomb potential in absence of off-diagonal terms. On the other hand, the SXX overestimates the dielectric screening, resulting in a slightly blue-shifted peak and a smaller oscillator strength for the first peak.
As was noted earlier \cite{Sun2020low}, gDDH gives the first excitonic peak lying between those by SXX and m-BSE. In fact, the gDDH spectrum is almost on top of the BSE,
showing an even better performance than tDDH. This disparity is attributed to the smaller RPA dielectric constant, which leads to a weaker screening and therefore a
stronger excitonic effect.

\begin{figure}
    \includegraphics[width=\linewidth]{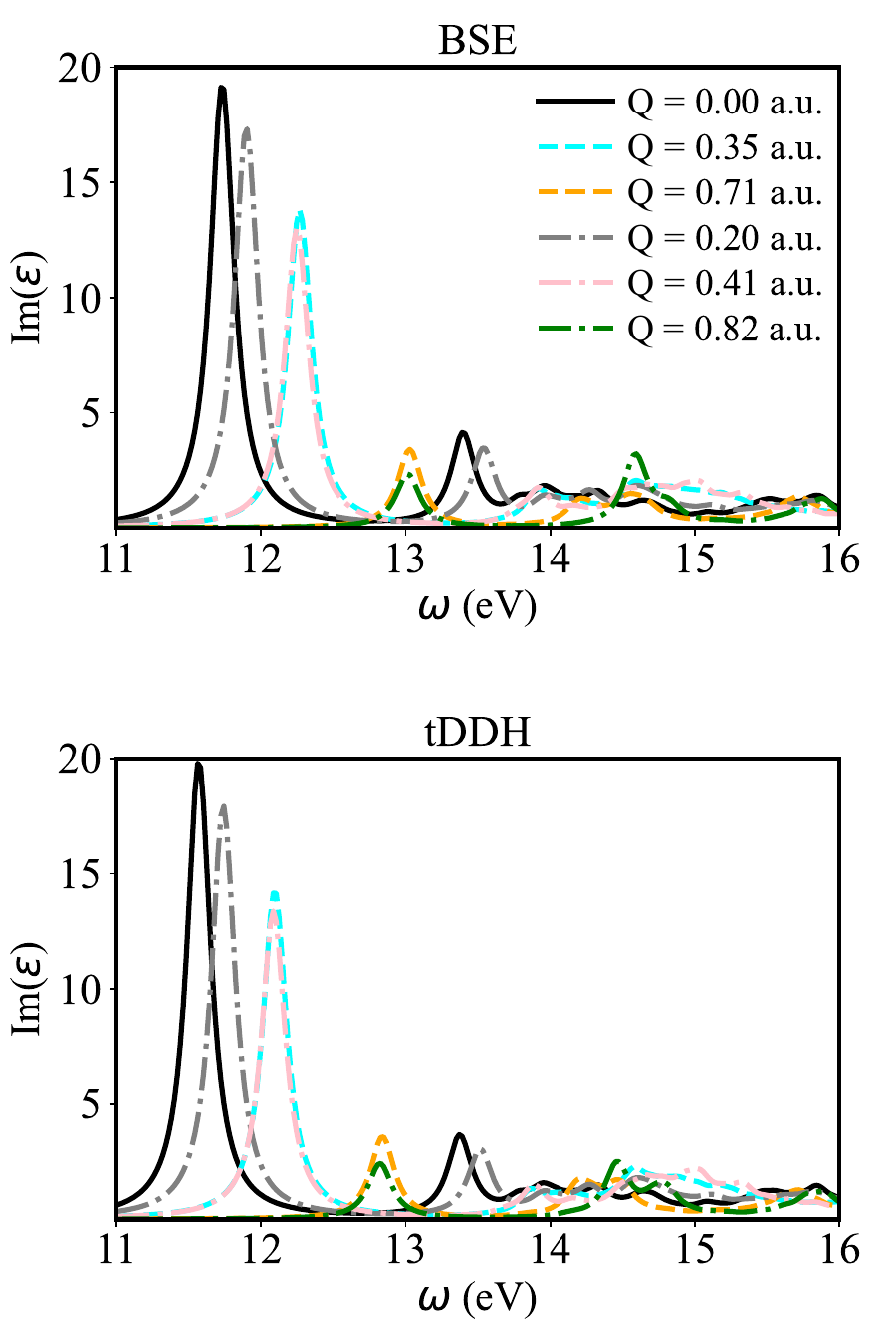}
    \caption{$\rm Im(\varepsilon)$ of LiF, calculated using BSE (top panel) and tDDH hybrid functional (bottom panel) at various momentum transfer values $Q$ along [111] (dashed lines) and [010] (dash-dotted lines) directions
    (see caption of Fig. \ref{fig:LiF_q} for the
    individual $\bfQ$-vectors).}
    \label{fig:LiF_compare}
\end{figure}

As the momentum transfer $Q$ increases, the excitonic peaks in the BSEs and hybrid functionals spectra start to shift toward higher energies and lose some of their oscillator strength, indicating a gradual weakening of excitonic binding energy as the transferred momentum increases. To see the shifting of the excitonic peaks more explicitly, Fig. \ref{fig:LiF_compare} superimposes $\rm Im(\varepsilon(\bfQ,\omega))$ for various momentum values, calculated with BSE and tDDH.
Clearly, even at higher values of $Q$ the spectra continue to show well-defined peaks, indicating that strong electron-hole interactions remain present, in agreement with observations by Gatti and Sottile \cite{gatti2013exciton}.
For this lowest excitonic peak, we found that all other methods maintain their performance compared to BSE when $\bfQ$ becomes finite.
However, a difference appears for the peaks in the range of 14-15 eV. Although the SXX underestimates the height of the first peak for all finite $\bfQ$, it gives the closest spectra in the range of 14-15 eV, not only for the position but also the height of the peaks. Nevertheless, our hybrid functionals are in very close agreement with the BSE over the entire range of momentum transfers considered, showing that they can successfully capture the strongly bound excitons and their trends as $\bfQ$ evolves.

\subsection{Comparison of CPU times}
Lastly, some remarks on computational efficiency. In our earlier work (see supplemental material of Ref. \cite{Sun2020low}) we had reported a significant computational speedup
of the SXX kernel compared to BSE, where the main benefit is due to the fact that SXX does not require calculating the full dielectric matrix.
We here confirm these observations for the case of LiF, focusing on tDDH with different recipes for the dielectric screening.

\begin{figure}
    \includegraphics[width=\linewidth]{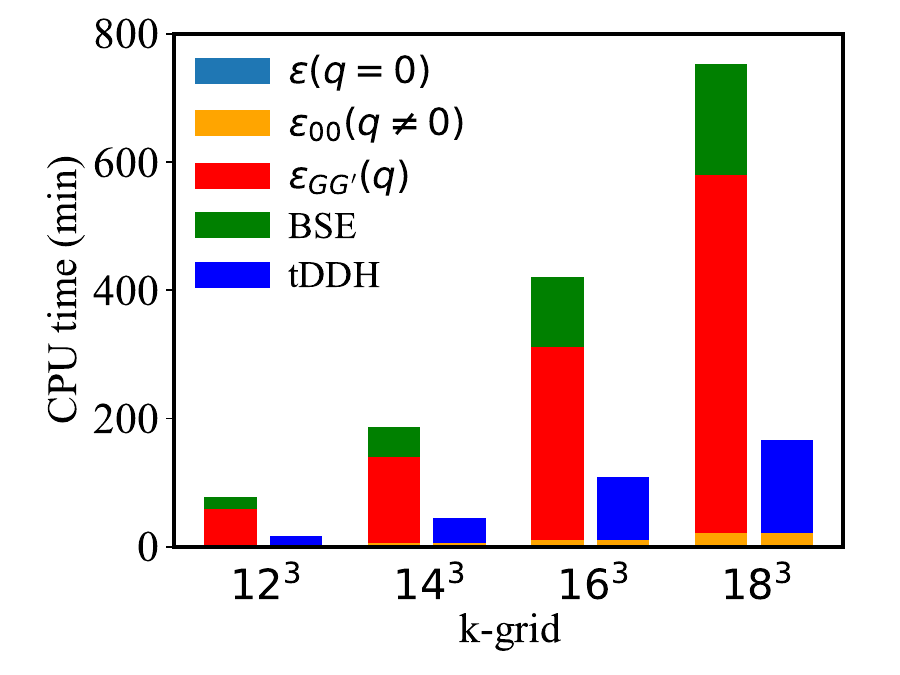}
    \caption{CPU times (in minutes) for calculating the dielectric function of LiF, comparing standard BSE and tDDH on top of different components of RPA dielectric screening, for various $k$-point grids.}
    \label{fig:cpu_time}
\end{figure}

Figure \ref{fig:cpu_time} compares BSE and hybrid calculations of $\rm Im(\varepsilon)$ on a single CPU with 32 cores, both using the same LDA+scissors band
structures as input. The total CPU time of standard BSE consists of the time for building the kernel and solving the BSE matrix equation, as well as the time for calculating the
full dielectric matrix $\varepsilon_{\bfG\bfG'}^{-1} (\mathbf{q})$ via RPA. The last part is not needed for the hybrids:
instead, calculating the macroscopic $\varepsilon(q=0)$ for gDDH or evaluating an empirical model $\varepsilon(q)$ for tDDH
requires negligible CPU times (invisible on the scale of Fig. \ref{fig:cpu_time}). Even a first-principles tDDH, which requires calculating the
$\Bar{\bfq}$-dependent $\varepsilon_{00}(\Bar{\bfq})$ head terms [see the discussion following Eq. (\ref{13})], costs much less than the effort to obtain the full $\varepsilon_{\bfG\bfG'}^{-1} (\mathbf{q})$. This can be seen in Fig. \ref{fig:cpu_time} by comparing the orange and red portions of the bars.

We also notice that the acceleration of the kernel part of the hybrids against the BSE is not as pronounced as that of m-BSE and SXX \cite{Sun2020low}, because
the ALDA kernel included in the hybrids adds extra cost for the off-diagonal terms of $f_{{\rm xc}, \bfG \bfG'}^{\rm ALDA}$ in Eq. (\ref{eq:12}). Nevertheless, a time saving of 10-15\% is observed for the kernel part of the hybrids against the BSE (compare the green and blue portions of the bars in Fig. Fig. \ref{fig:cpu_time}), indicating that
neglecting the off-diagonal terms in the direct electron-hole interaction [$K^{\rm SXX}$ in Eq.~(\ref{eq:11})] is the dominant effect.
In addition, our further tests with only 1 core  (not shown here) even indicate that the hybrids kernel would be 10 times faster than BSE kernel, which is not the real case in current research but instructive somehow.

Moreover, the overall hybrid calculations become increasingly faster compared to BSE with increasing $k$-point grid size. The main increment of time saving comes from the approximation of RPA calculations for dielectric screening, which demonstrates the advantage of the truncated dielectric screening scheme.

It goes without saying that the required CPU time of a computation depends on many details and parameters, and
can vary widely for different implementations. The absolute values of the CPU times reported here are certainly
not meant to be viewed as performance benchmarks; the overall trends and the differences between the two methods, however,
are likely to persist for other platforms and implementations.

\section{Discussions and Conclusions} \label{Summary}
In this paper we have investigated the momentum dependence of the dielectric function in materials using a dielectrically screened hybrid functional approach. The study focused on the accurate calculation of excitonic features in $\varepsilon(\bfQ,\omega)$ in three representative materials (Si, GaN and LiF), which are known to exhibit different electronic and optical properties. By combining the strengths of advanced computational methods like BSE and hybrid functionals, this paper aims to provide an improved understanding of exciton behavior in momentum space, which is crucial for next-generation optoelectronic and quantum information devices.

One of the key results of this study is the demonstration that our dielectrically screened hybrid functionals closely reproduce BSE results over a wide range of momentum transfers, thus providing new insights into exciton dispersions in momentum space. At large momentum transfer, we find for Si and GaN that the hybrid functional behaves similarly to the ALDA, which is known to provide a good description of the short-range many-body effects that are required for
$\rm Im(\varepsilon)$ at finite $\bfQ$ in these systems. Therefore, generalized TDDFT using specialized hybrid kernels is emerging as a promising and cost-effective alternative to full many-body perturbation (BSE) calculations.

We also investigated the performance of the range-separated screened hybrid functional with screening wavevectors confined in the first Brillouin zone (tDDH), similar to the m-BSE approach, and found that it yields results very close to the BSE over the entire range of $\bfQ$ values
considered here. In fact, tDDH turned out to be significantly closer to BSE than the gDDH approach. However, past studies \cite{Weissker2010,sharma2012enhanced} have shown
that the ALDA often provides an excellent description of IXS and EELS at finite momentum transfer, reproducing some spectral parts better than the BSE -- and the gDDH functional
becomes more and more similar to the ALDA for large momenta, in contrast with tDDH.
Thus, which hybrid kernel is to be preferred will certainly depend on the property to be simulated, and will require detailed
assessment using experimental results; this will be a task for future investigations.

We emphasize that the hybrid approach presented here is appropriate for 3D bulk materials, where the dielectric function can already be successfully approximated
by just one parameter $\gamma = \varepsilon_{00}(\bfq=0)$. In two dimensions (2D), screening is more complex, with a dielectric function that
has a small-$q$ behavior of the form $\varepsilon(q,0) = 1 + \alpha q$ \cite{Cudazzo2011,Thygesen2017,Trolle2017}.
Given the current interest in 2D materials and their applications \cite{Hulser2013,Qiu2013,Qiu2016}, especially due to their strong excitonic effects,
extending the dielectrically screened TDDFT hybrid approach to 2D is an important and currently ongoing task.

\acknowledgments
D. A. and C. A. U.  acknowledge support by NSF Grant No. DMR-2149082. J. S.  acknowledges support by the National Natural Science Foundation of China (No.22303041 and No. 12274228), the NSF of Jiangsu Province (No. BK20230908) and Fundamental Research Funds for the Central Universities (No. 30923010203). The computations for this work were performed on the high-performance computing infrastructure provided by Research Computing Support Services at the University of Missouri-Columbia.

\bibliography{paper}

\end{document}